*Article*

# Exploring AI Tool's Versatile Responses: An In-depth Analysis Across Different Industries and Its Performance Evaluation

**Hitesh Mohapatra** , **Soumya Ranjan Mishra**

1*,2 School of Computer Engineering, KIIT (Deemed to be) University, Bhubaneswar - 751024, Odisha, India
* Correspondence: hiteshmahapatra@gmail.com; soumyaranjanmishra.in@gmail.com

**Abstract:** AI Tool is a large language model (LLM) designed to generate human-like responses in natural language conversations. It is trained on a massive corpus of text from the internet, which allows it to leverage a broad understanding of language, general knowledge, and various domains. AI Tool can provide information, engage in conversations, assist with tasks, and even offer creative suggestions. The underlying technology behind AI Tool is a transformer neural network. Transformers excel at capturing long-range dependencies in text, making them well-suited for language-related tasks. AI Tool has 175 billion parameters, making it one of the largest and most powerful LLMs to date. This work presents an overview of AI Tool's responses on various sectors of industry. Further, the responses of AI Tool have been cross-verified with human experts in the corresponding fields. To validate the performance of AI Tool, a few explicit parameters have been considered and the evaluation has been done. This study will help the research community and other users to understand the uses of AI Tool and its interaction pattern. The results of this study show that AI Tool is able to generate human-like responses that are both informative and engaging. However, it is important to note that AI Tool can occasionally produce incorrect or nonsensical answers. It is therefore important to critically evaluate the information that AI Tool provides and to verify it from reliable sources when necessary. Overall, this study suggests that AI Tool is a promising new tool for natural language processing, and that it has the potential to be used in a wide variety of applications.

**Keywords:** AI Tool, Response Pattern, Societal impact, Chatbots, Response evaluation, performance metric

## 1. Introduction

AI Tool is an advanced conversational AI model developed by OpenAI. It is designed to engage in natural language conversations with users, offering responses that are coherent and contextually relevant. AI Tool builds upon the success of previous iterations, such as GPT-2 and GPT-3, incorporating improvements in training methodologies and model architecture. The model utilizes a transformer neural network, which is a deep learning architecture that excels at processing sequential data, such as text. Transformers allow AI Tool to capture long-range dependencies and understand the context of a conversation, making it capable of generating human-like responses. AI Tool has been trained on a massive data-set of text and code, which allows it to have a broad understanding of language and knowledge. This allows AI Tool to respond to a wide range of prompts and questions, including those that are open ended, challenging, or strange. AI Tool is still under development, but it has the potential to be a powerful tool for a variety of applications, such as customer service, education, and research [1]. Table.1 illustrates the evolution of AI chatbots with their properties.

AI Tool has been trained on a massive corpus of text from the internet, encompassing a wide range of topics and domains. This training data enables the model to have a broad understanding of language, facts, and cultural knowledge. However, it is important to note that AI Tool does not possess real-time information and its knowledge is based on data available up until September 2021 [2]. OpenAI has made efforts to ensure that AI

**Table 1.** ChatGPT Versions and Comparision

| Version | Uses | Architecture | Parameter count | Year |
|---------|------|--------------|-----------------|------|
| GPT-1 | General | 12-layer, 12-headed Transformer decoder (without encoder), followed by linear-softmax, trained on Book Corpus with a dataset size of 4.5GB of text | 117 million | 2018 |
| GPT-2 | General | Similar to GPT-1, but with adjusted normalization techniques, trained on Web Text dataset consisting of 40GB of text | 1.5 billion | 2019 |
| GPT-3 | General | An extension of GPT-2, incorporating alterations to enable greater scalability, trained on a dataset of 570 GB plaintext | 175 billion | 2020 |
| InstructGPT | Conversation | GPT-3 fine-tuned through a human feedback model to enhance its ability to comprehend and follow instructions | 175 billion | 2022 |
| ProtGPT2 | Protein Sequences | Modeled similar to GPT-2 large (36 layers), utilizing Protein sequences sourced from UniRef50, totaling 44.88 million sequences | 738 million | 2022 |
| BioGPT | Biomedical Content | Following the framework of GPT-2 medium (24 layers, 16 heads), incorporating non-empty items extracted from a PubMed dataset, totaling 1.5 million | 347 million | 2022 |
| ChatGPT | Dialogue | Built upon GPT-3.5, and refined through a combination of supervised learning and reinforcement learning with input from human feedback (RLHF) | 175 billion | 2022 |
| GPT-4 | General | Trained through a dual approach involving text prediction and RLHF, capable of accepting both textual and image inputs, including third-party data | 100 trillion | 2023 |

Tool exhibits responsible behavior by minimizing biased and offensive outputs. During training, the model is fine-tuned and guided using a combination of human reviewers and reinforcement learning algorithms to align its responses with desired ethical standards. Despite these efforts, the model may occasionally produce incorrect, nonsensical, or biased responses [3].

AI Tool is a large language model (LLM) that can be used for a variety of purposes, including answering questions, providing explanations, assisting with tasks, generating ideas, and engaging in creative writing [4]. It has found applications in customer support, content generation, language learning, brainstorming, and more. OpenAI has made AI Tool accessible through various interfaces, including web-based platforms and API services. This allows developers and users to interact with the model and integrate it into their own applications or services [5].

## 1.1. Working of AI Tool

The transformer model consists of layers of self-attention mechanisms and feed-forward neural networks, enabling it to capture complex patterns and dependencies in language. Here's a simplified overview of how AI Tool works is illustrated in Figure.1. The first step is tokenization where the input text is divided into tokens, which can be as small as individual characters or as large as whole words. Each token is assigned a numerical representation. The second step is encoding where the tokens are passed through an initial embedding layer, where they are transformed into high-dimensional vectors that capture their semantic meaning. The third step is self-Attention where the encoded tokens are then processed through multiple layers of self-attention mechanisms. Self-attention allows the model to weigh the importance of each token based on its relationship with other tokens in the input sequence. This helps the model understand the context and dependencies within the text. The fourth step is feed-forward networks where after the self-attention layers, the output is passed through a series of feed-forward neural networks. These networks apply non-linear transformations to the token representations, further capturing complex patterns in the data.

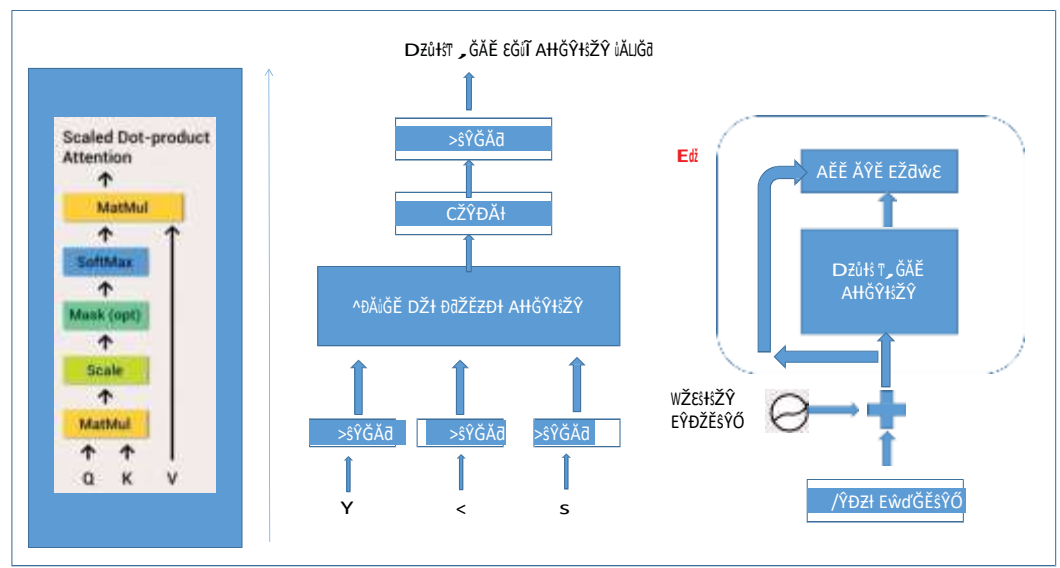

**Figure 1.** Architecture of AI-Tool

During the decoding phase the final output of the feed-forward networks is passed through a decoding layer, which maps the representations back to the vocabulary space. This allows the model to generate the next token or predict the most likely token given the context. After that during response generation the conversation happens where, AI Tool takes a user's input and generates a response based on the learned patterns and dependencies in the training data. The model generates tokens one by one, taking into account the preceding context and user input. This process continues until an appropriate response is generated or a maximum response length is reached. Training AI Tool involves a two-step process. Such as

1. **Pre-training:** The model is trained on a large corpus of text from the internet. By predicting the next token in a sentence, the model learns to understand language and capture various patterns and concepts. This pre-training allows AI Tool to acquire a broad knowledge of language and facts.

2. **Fine-tuning:** After pre-training, the model is fine-tuned using a more specific dataset, which includes demonstrations and comparisons by human reviewers. OpenAI provides guidelines to the reviewers to ensure the model's responses align with desired behavior and ethical standards. This fine-tuning helps shape the model's behavior and allows it to produce more appropriate and controlled responses.

The main idea behind the transformer is to capture the relationships between different words or tokens in a sentence, allowing for better context understanding and generation. The transformer model consists of two main components: the encoder and the decoder. In the context of Chat boat, the encoder processes the input message and the conversation history, while the decoder generates the response. The encoder and decoder are composed of multiple layers, with each layer containing sub-modules. Let's focus on one layer to understand the core workings of the Transformer.

Self-Attention Mechanism: Self-attention is a mechanism that allows the model to weigh the importance of each word/token in the input sequence. For each word/token, self-attention computes its "attention scores" with respect to all other words/tokens in the sequence. These attention scores determine how much focus the model should place on each word/token during processing. The attention scores are computed using three learned matrices: Query, Key, and Value. These are multiplied together to produce the attention scores. Multi-Head Attention: To capture different types of dependencies and improve performance, self-attention is applied in parallel multiple times, known as "heads." Each attention head learns different relationships between words/tokens, allowing the model to attend to various aspects of the input. The outputs of all attention heads are concatenated and linearly transformed to retain relevant information.

Position-wise Feed-Forward Networks: After self-attention, the output is passed through a position-wise feed-forward network. This network consists of two linear layers with a non-linear activation function in between, allowing the model to transform and combine information across positions. Residual Connections and Layer Normalization: To address the challenge of vanishing gradients, residual connections are added, allowing the model to retain information from previous layers. Layer normalization is applied after each sub-module, ensuring stable gradients during training. The encoder processes the input message and conversation history by stacking multiple layers of self-attention and feed-forward networks. The decoder, on the other hand, also incorporates an additional attention mechanism that attends over the encoder's output to capture relevant context information. During training, the model is optimized to generate coherent and contextually appropriate responses using techniques such as maximum likelihood estimation. The parameters of the model are learned by minimizing the discrepancy between the model's generated responses and the ground truth responses in the training data.

The paper's structure is outlined below. Section 2 provides an in-depth examination of the related literature. In Section 3, we delve into our exploration using the AI tool. Technical domains and specific applications are thoroughly discussed in Section 4, while Section 5 similarly addresses business and administrative sectors along with pertinent applications. Section 6 provides a concise overview of observations and performance evaluations with behavioral analysis of the AI tool followed by the conclusion and references.

## 2. Related Work

AI Tool has gained significant attention worldwide and sparked discussions in academia. Some individuals foresee potential disruptions, such as the decline of assigned essays and widespread unemployment as machines assume writing tasks. There are concerns regarding the prevalence of cheating, with worries that it may become challenging or even impossible to detect. Moreover, fears persist that students may become complacent, convinced that they can rely entirely on automated writing. This article aims to provide a concise and accurate overview of AI Tool, including its capabilities and limitations. Additionally, it explores methods for identifying AI-driven cheating, as well as potential applications that could alleviate workload, enhance students' writing skills, facilitate exam composition and grading, and elevate the quality of research papers [6].

In this commentary, we delve into the subject of AI Tool and offer our insights regarding its potential usefulness in systematic reviews (SRs). We assess the appropriateness and applicability of AI Tool's responses to prompts related to SRs. The rise of artificial intelligence (AI)-assisted technologies has prompted discussions on their current capabilities, limitations, and opportunities for integration into scientific endeavors. Large language models (LLMs), such as OpenAI's AI Tool, have garnered significant attention due to their ability to generate natural-sounding responses across various prompts. SRs, which rely on secondary data and often demand substantial time and financial resources, present an attractive domain for the development of AI-assistive technologies. On February 6, 2023, the PICO Portal developers conducted a webinar to explore how AI Tool responds to tasks associated with SR methodology. Based on our exploration of AI Tool's responses, we observe that while AI Tool and LLMs show promise in assisting with SR-related tasks, the technology is still in its early stages and requires significant further development for such applications [7].

The application of AI Tool can be used in multiple discipline of science. The aim was to explore the application of AI in various tasks within the mechanical engineering domain and draw conclusions through statistical analysis of the outcomes. However, when utilizing AI Tool in several calculation examples, we discovered that it generated incorrect results, erroneous formulas, and similar inaccuracies [8]. The firing appropriate query can help to use the AI Tool in software development and modelling too. AI Tool has demonstrated its capability to generate useful code snippets that, on occasion, are correct and successfully accomplish the desired task specified by the user. Moreover, AI Tool exhibits familiarity with various textual modeling languages, including domain-specific languages (DSLs). Notably, the example of GraphGPT showcases the potential for language designers to instruct AI Tool on the desired structure of a modeling language, resulting in the generation of code fragments within that language. In the case of GraphGPT, it employs a clever approach of requesting a JSON-encoded representation of the graph, which can then be rendered into a diagram. The possibilities for leveraging generative AI in modeling are vast, and the exact ways in which it will transform the business and practices of modeling in the future remain uncertain[9]. The uses of AI Tool can also be found in scientific abstract writing [10] and in health sector [11] too.

In the field of academic writing AI Tool can be used in revolutionary way [12]. As AI Tool is a new tool hence many researchers is trying to explore it in a several way. One such method is chat with AI Tool where the authors have shared their experience [13]. Though AI Tool is a powerful tool that can be used in many applications but there are several instance where the faulty references can be found [14]. Since its launch in 2022, AI Tool, a query-oriented language generation tool, has garnered significant attention. Although the initial excitement may have waned, the impact of AI Tool has sparked lasting structural changes. Notably, academic journals have published papers with AI Tool listed as an author, while certain educational institutions have opted to prohibit its use due to concerns about potential misuse. Criticisms of AI Tool have primarily revolved around its inaccuracies, often labeling it as a "bullshit generator." Additionally, some have highlighted the undesirable consequences that arise from its utilization, such as the potential to undermine creativity. However, we contend that there is an unaddressed issue at hand — the fundamental ideas and politics that drive the development of these tools and facilitate their uncritical adoption [15].

## 3. Exploration with AI Tool

The scope of AI tools is vast and extends across diverse sectors, revolutionizing the way we approach challenges and opportunities. From healthcare and finance to manufacturing and entertainment, AI's transformative capabilities have left an indelible mark. In healthcare, AI aids in diagnosing diseases and personalizing treatment plans, while in finance, it enhances fraud detection and market analysis. Industries like manufacturing benefit from AI-driven automation for improved efficiency, and the entertainment sector leverages AI to create immersive experiences. The ability of AI tools to analyze vast data-sets, recognize patterns, and make informed decisions transcends boundaries, making them an indispensable asset in shaping the present and future of countless sectors [16]. The study has focused on two types of sector classification such as technical and business administrative sectors respectively. Further, each type of category have considered 5 different types of individual sectors for response analysis.

Technical Sectors

- Medical and Health Case Sector
- Software Development
- Smart Agriculture
- Logistics and Supply
- Smart City Designing

Business Administrative Sectors

- Education and Academic
- Crime Monitoring
- Administrative Accounting
- Entertainment Industry
- Culture and Value Promotion

## 4. AI Tool Response in Technical Sectors

### 4.1. Medical and Health Sector

AI tools have a transformative role in the medical and health sector. They serve as conversational interfaces for quick access to medical information, including research papers, clinical guidelines, and drug details. Patient education is enhanced through personalized health insights and answers to queries [4]. AI aids in symptom assessment and initial guidance for seeking medical help. In telemedicine, AI integrates for remote patient monitoring and virtual consultations, ensuring data gathering and medication reminders. Analyzing electronic health records, AI organizes patient data for efficient decision-making [5]. It supports mental health by offering coping strategies and resources. Medical education benefits from AI simulations and feedback. In research, AI extracts information, aids in data analysis, and facilitates patient recruitment for clinical trials. The generated response from AI Tool has been cross verified with opinions of experts medical and health sector. For the evaluation process we have communicated with 34 doctors and 16 health care sector staffs. Figure.2a illustrates the performance evaluation based on the parameters that are considered in Table.1.

### 4.2. Software Development

AI Tool's role in software development complements but doesn't replace human expertise. While it aids coding tasks, debugging, and documentation, human validation remains vital. AI generates code snippets, suggests functions,  and completes segments. It assists in debugging by proposing solutions to errors. AI simplifies documentation creation, extracting insights from source code. In applications, it enables natural language interactions. AI suggests code improvements and best practices. It acts as a knowledge hub, explaining concepts and offering examples. Moreover, AI enhances teamwork through project management assistance.

Further, AI tool can aid in test case generation by analyzing requirements or specifications and suggesting relevant test scenarios or edge cases. It can assist in ensuring thorough test coverage and identifying potential issues. AI Tool can provide guidance on version control systems like Git. It can assist developers in understanding branching strategies, resolving merge conflicts, and recommending best practices for collaboration and code management. AI Tool can offer guidance on setting up development environments, configuring tools, and resolving environment-related issues. It can assist developers in getting started with specific frameworks or platforms. The generated response from AI Tool has been cross verified with opinions of developers and testers. For the evaluation process we have communicated with 30 software developers and 21 testing engineers. Figure. 2b presents the performance evaluation based on the parameters that are considered in Table.1.

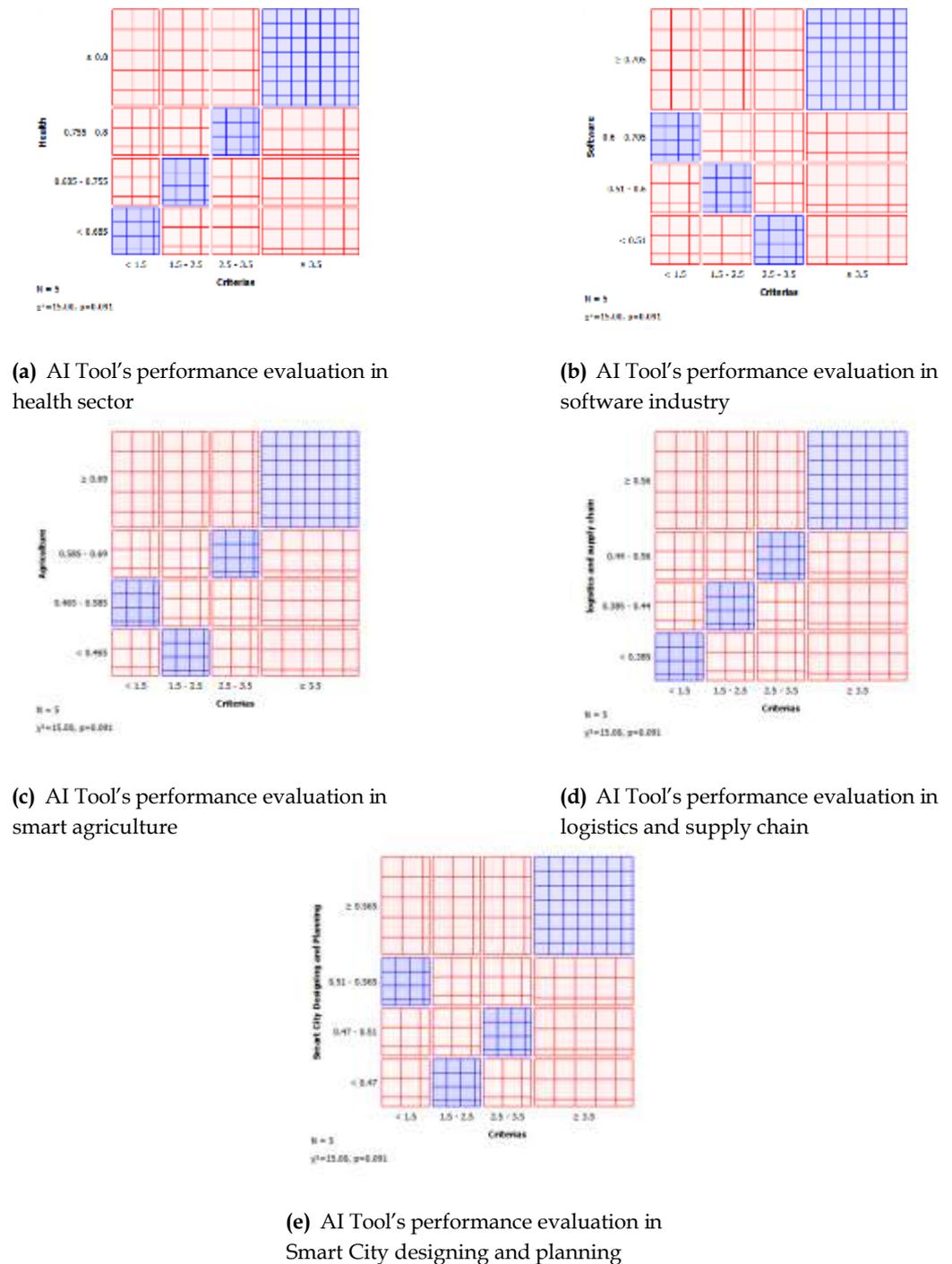

**(a)** AI Tool's performance evaluation in health sector

**(b)** AI Tool's performance evaluation in software industry

**(c)** AI Tool's performance evaluation in smart agriculture

**(d)** AI Tool's performance evaluation in logistics and supply chain

**(e)** AI Tool's performance evaluation in Smart City designing and planning

**Figure 2.** AI Tool's performance evaluation in different technical sectors

### 4.3. Smart agriculture

Leveraging AI Tool in smart agriculture boosts productivity, resource efficiency, and sustainability. It offers real-time insights for crop management, utilizing weather, soil, and crop data. AI detects pests and diseases early, suggesting solutions. It optimizes resource use in precision farming by analyzing sensor and historical data, enhancing efficiency. AI Tool benefits smart agriculture through weather updates, aiding decisions on planting and protective measures. It advises crop choices, rotation, and diversification. Soil health monitoring and improvement recommendations are provided. Efficient water management guidance is given using sensor and forecast data. AI Tool offers market insights, aiding crop selection and profitability. It acts as a virtual advisor, answering farming queries and providing educational resources. The generated response from AI Tool has been cross verified with opinions of farmers of both conventional and smart agriculture. For the evaluation process we have communicated with 23 farmers and 21 researchers of agricultural science. Figure.2c presents the performance evaluation based on the parameters that are considered in Table.1.

### 4.4. Logistic and supply chain management

While AI Tool aids logistic and supply chain management, human expertise remains vital. AI serves as a virtual assistant, handling customer queries and providing real-time support. It integrates with systems for personalized order tracking. AI analyzes data for inventory, demand, and production, offering supply chain optimization suggestions. However, human oversight is essential for complex situations and critical decisions based on AI recommendations. AI enhances efficiency but should be employed in tandem with human judgment [17]. AI Tool streamlines supplier management, handling routine inquiries and supplier performance insights. It identifies new suppliers and aids communication. AI analyzes supply chain data, predicting demand changes and evaluating external risks. It suggests contingency plans, enhancing decision-making. AI acts as a training tool, simulating scenarios for risk-free practice. It shares knowledge on best practices and emerging trends. AI enhances supplier collaboration, risk assessment, and skill development in supply chain management. The generated response from AI Tool has been cross verified with opinions of business holders and logistic department of various suppliers. For the evaluation process we have communicated with 13 businessmen and 9 logistic suppliers. Figure.2d presents the performance evaluation based on the parameters that are considered in Table.1.

### 4.5. Smart city designing and planning

AI Tool enhances smart city planning with data-driven insights and citizen engagement. It aids decision-making, creating livable urban environments. AI acts as a virtual assistant, gathering citizen input and feedback [18]. It analyzes intricate urban data for trends and visualizations, aiding informed city planning. AI Tool contributes to urban planning by suggesting designs based on population, transport, green spaces, and energy [19]. It optimizes land use and connectivity while integrating sustainability. AI analyzes real-time traffic data, optimizing flow and reducing congestion [20]. It recommends traffic management systems and efficient routes. AI aids in energy management strategies for smart cities [21]. AI Tool enhances urban planning by analyzing energy patterns, suggesting efficient technologies, and integrating renewables. It optimizes energy distribution, monitors air quality, and manages waste. AI aids in emergency planning, analyzing data for disaster preparedness and resource allocation. It fosters stakeholder collaboration, acting as a knowledge base. AI integrates sustainability and green initiatives, promoting eco-friendly tech and recycling programs. It evaluates policy impact on smart cities, simulating effects on transportation, energy, and services to inform decisions [22].

The generated response from AI Tool has been cross verified with opinions of city planners and civil engineers. For the evaluation process we have communicated with 4 city planners and 7 civil engineers. Figure.2e presents the performance evaluation based on the parameters that are considered in Table.1.

## 5. AI Tool Response in Business Administrative Sectors

### 5.1. Education and Academic Paper Writing

Maintaining a balanced approach to AI Tool integration in education is crucial, emphasizing human guidance for fostering critical thinking [3]. AI acts as a virtual tutor, delivering personalized guidance and learning resources [23]. It aids academic writing with grammar suggestions and feedback [24]. AI assists in research by retrieving articles, summarizing papers, and detecting plagiarism [25]. It supports language learners through practice exercises and explanations. Additionally, AI aids researchers by generating initial drafts and structuring papers [26]. AI Tool can assist in the peer review process by analyzing submitted manuscripts, identifying potential issues, and offering constructive feedback. It can help reviewers focus on important aspects such as clarity, methodology, and validity of research. AI Tool can aid in conducting literature reviews by extracting relevant information from academic articles, summarizing key findings, and organizing references. It can save time for researchers in the initial stages of their literature review process. AI Tool can provide information and resources on academic integrity and ethical writing practices. It can educate students about citation rules, paraphrasing techniques, and the importance of avoiding plagiarism. The generated response from AI Tool has been cross verified with opinions of professors and students. For the evaluation process we have communicated with 30 professors and 60 students. Figure.3a presents the performance evaluation based on the parameters that are considered in Table.1.

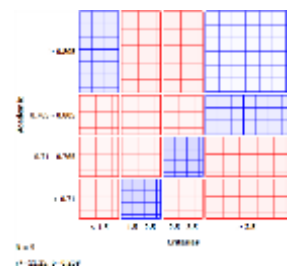

**(a)** AI Tool's performance evaluation in academic sector

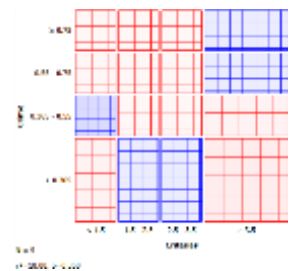

**(b)** AI Tool's performance evaluation in crime monitoring

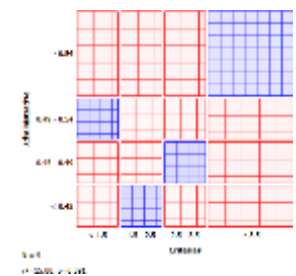

**(c)** AI Tool's performance evaluation in administrative section

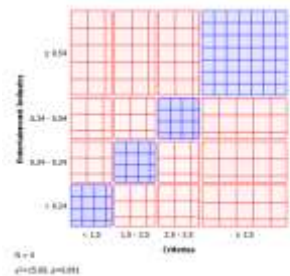

**(d)** AI Tool's performance evaluation in entertainment industry

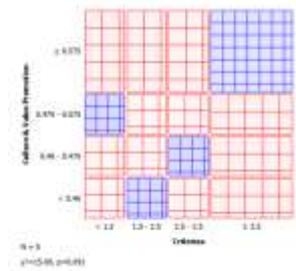

**(e)** AI Tool's performance evaluation in culture and value promotion

**Figure 3.** AI Tool's performance evaluation in different business administrative sectors

### 5.2. Crime monitoring

AI supports law enforcement but doesn't replace human decision-making. It interfaces for incident reporting, enhancing public communication. Analyzing crime data, AI identifies patterns and trends for resource allocation. It predicts crime probabilities, enabling proactive measures by law enforcement in high-risk areas. AI Tool assists in suspect identification using witness descriptions or images. It generates composite sketches and matches from databases. It aids investigators with data retrieval from public records, social media, and databases for background checks and connections. AI educates the public on safety and crime prevention. It analyzes OSINT data for threat detection and crime activities. AI translates languages for communication and analyzes text data for threat identification. The generated response from AI Tool has been cross verified with opinions of people of judicial. For the evaluation process we have communicated with 9 police officers and 11 lawyers. Figure.3b presents the performance evaluation based on the parameters that are considered in Table.1.

### 5.3. Administrative Actions

Leveraging AI Tool streamlines administrative actions, allowing focus on strategic tasks [27] where balancing automation and human touch is the key. AI serves as virtual support, addressing inquiries and simple issues 24/7 [17]. It schedules appointments and manages calendars efficiently. AI retrieves data for administrators, aiding quick decision-making. AI Tool simplifies form filling and document completion. It assists new employee on-boarding, sharing policies and guidance. AI offers information on policies, procedures, and compliance [28]. It analyzes data, generates user-friendly reports for informed decisions [29]. AI acts as a virtual trainer, aiding ongoing learning. It automates routine tasks, integrates with software systems for efficiency. AI sends notifications and reminders, ensuring effective communication. The generated response from AI Tool has been cross verified with opinions of 34 clerical staffs of administrative section at KIIT University. Figure.3c presents the performance evaluation based on the parameters that are considered in Table.1.

### 5.4. Entertainment Industry

AI Tool empowers the entertainment industry with innovative content creation and interactive experiences. It generates scripts, dialogues, and characters, fostering creativity. AI enhances storytelling, character development, and narrative exploration. It crafts interactive virtual characters, enabling immersive experiences across various platforms. AI Tool transforms entertainment with personalized recommendations, interactive storytelling, and virtual characters. It analyzes preferences for suggestions, creating immersive experiences. AI shapes narratives, responds to choices, and offers personalized storylines. It crafts virtual assistants for celebs or characters, deepening audience connections. In video games, AI enhances dialogues and character depth. It engages fans on social media, maintaining interactive presence. AI elevates live events with real-time engagement and interactive elements. It generates synthetic voices for various roles. AI sparks fan engagement by simulating conversations and discussing fictional worlds, fostering creativity and community [30]. The generated response from AI Tool has been cross verified with opinions of content creators on YouTube. For the evaluation process we have communicated with 34 content creators on YouTube. Figure.3d presents the performance evaluation based on the parameters that are considered in Table.1.

### 5.5. Culture and value promotion

AI Tool advances cultural understanding and inclusivity. It educates about diverse cultures, traditions, and languages. It acts as a virtual guide, offering insights into history, art, and practices. AI aids language learning, facilitating cross-cultural communication. It fosters empathy and appreciation for global cultures, contributing to a more culturally rich society[31]. AI Tool enables virtual cultural exchange, connecting individuals globally. It discusses art forms, recommends artists, and sparks creativity. AI engages in ethical discussions, fostering dialogue on values and morals. It simulates cultural exhibitions, providing historical context and interactivity. AI preserves heritage by organizing cultural information digitally. It fosters intercultural dialogue, artistic appreciation, and ethical reflection, enriching global connections and cultural understanding[32]. AI Tool aids cultural tourism, suggesting landmarks and events. It fosters discussions on values and decisions, enhancing values-based choices. AI supports social impact campaigns for diversity and inclusivity, interacting with users to raise awareness and encourage actions. It promotes cultural engagement, values-based decisions, and positive impact, contributing to a more informed and culturally aware society[33]. The generated response from AI Tool has been cross verified with opinions of eleven humanities and social science professors of KIIT University. Figure.3e presents the performance evaluation based on the parameters that are considered in Table.1.

## 6. Observation on AI Tool behaviour

Though AI Tool become a buzz word and in the last two sections we have witnessed the implications of AI Tool on various sectors of the society. Though it is powerful tool for the modern digital world still it has many negative and positive side effects.

### 6.1. Benefits of using AI Tool

These benefits demonstrate the potential of AI Tool to enhance communication, streamline processes, improve user experiences, and provide valuable support across various domains. However, it is essential to consider and address potential limitations, ethical considerations, and user expectations to effectively harness these benefits.

- Quick and convenient communication: AI Tool can provide instant responses to user queries, eliminating the need for waiting or queuing. This can improve customer satisfaction and reduce the time it takes to resolve issues.
- Round-the-clock availability: AI Tool can operate 24/7, providing support and information at any time. This can be especially helpful for businesses that operate in multiple time zones or that have a global customer base.
- Cost-effectiveness: AI Tool can be cost-effective compared to maintaining a large customer support team or hiring additional staff. Once deployed, it can handle multiple conversations simultaneously, reducing the need for human resources and lowering operational costs.
- Scalability: AI Tool can handle a high volume of conversations simultaneously, making it highly scalable. As the user base grows, the system can accommodate increased demand without significant infrastructure or resource investments.
- Consistency and accuracy: AI Tool can provide consistent and accurate responses based on the training it has received. This can avoid human errors and inconsistencies that may arise from manual interactions, ensuring a high level of reliability and accuracy.
- Multilingual support: AI Tool can support multiple languages, enabling communication with users from diverse linguistic backgrounds. This can be especially helpful for businesses that operate in a global marketplace.
- Interactive and dynamic conversations: AI Tool can engage users in interactive and dynamic conversations, providing a personalized and tailored experience. This can lead to increased user engagement and satisfaction.
- Knowledge repository: AI Tool can act as a knowledge repository, storing and retrieving information on a wide range of topics. This can help users to find the information they need quickly and easily.
- Continuous learning: AI Tool can continuously learn from user interactions, improving its responses and performance over time. This can lead to a better overall user experience.
- Augmentation of human capabilities: AI Tool can augment human capabilities by assisting with information retrieval, decision-making, and task automation. This can free up human operators to focus on more complex or specialized tasks.

### 6.2. Side-effects of blind using of AI Tool

Here, we have listed some of the primary negative impacts. To mitigate these negative impacts, responsible development, transparent practices, ongoing research, and regulation are important. Ethical guidelines, bias detection, and correction mechanisms, as well as user education on the limitations and potential biases of AI systems, can help address these concerns and promote the responsible use of AI Tool and similar technologies. While AI Tool and similar language models offer many benefits, there are also potential negative impacts on society.

- Misinformation and propaganda: AI Tool can generate text based on the input it receives, including false or misleading information. If used irresponsibly or without proper oversight, it can contribute to the spread of misinformation, conspiracy theories, or propaganda, which can undermine trust, create confusion, and harm society.
- Bias: AI Tool learns from the data it's trained on, which can include biases present in the training data. If the training data contains biases related to race, gender, or other sensitive topics, AI Tool may inadvertently perpetuate and amplify these biases in its generated responses, leading to unfair or discriminatory outcomes.
- Accuracy and reliability: AI Tool operates as a tool, and its outputs are not independently verified or fact-checked. This lack of accountability raises concerns about the accuracy and reliability of the information it provides, potentially leading users to make decisions based on flawed or misleading advice.
- Ethical exploitation: AI Tool can be exploited for unethical purposes, such as generating malicious content, engaging in harmful behaviors, or deceiving individuals. This raises concerns about privacy, security, and the potential for abuse by malicious actors.
- Reduced human-to-human interaction: Over-reliance on AI Tool for communication and problem-solving can reduce human-to-human interaction, which is crucial for building social connections, empathy, and emotional intelligence. Excessive dependence on AI-powered systems may lead to a decline in interpersonal skills and hinder the development of authentic relationships.
- Job displacement: The increased automation and efficiency offered by AI Tool may lead to job displacement in certain industries. Tasks that were previously performed by humans, such as customer support or content generation, may be taken over by AI systems, potentially resulting in unemployment or the need for reskilling.

AI Tool can be used for social engineering or manipulating individuals by generating persuasive or deceptive messages. Malicious actors could exploit this technology to deceive or exploit unsuspecting users for personal gain or malicious purposes. Interacting with AI systems like AI Tool may have psychological effects on individuals. For some users, relying heavily on AI-generated responses for emotional support or guidance could lead to a sense of detachment, isolation, or a reduction in critical thinking skills. AI Tool interacts with users and collects data during conversations, raising privacy concerns. Depending on the use and storage of this data, there is a risk of misuse, unauthorized access, or breaches that compromise user privacy and security. The widespread adoption and benefits of AI Tool may not be accessible to all individuals due to factors such as cost, infrastructure limitations, or digital literacy. This can contribute to a technological divide, exacerbating existing inequalities in society.

### 6.3. Numerical performance evaluation of AI Tool responses

By employing these approaches, you can gain insights into the performance of AI Tool, understand its strengths and limitations, and make informed decisions to improve its overall performance and user satisfaction. Analyzing the performance of AI Tool responses can help assess its effectiveness and identify areas for improvement. Here are some approaches to analyze its performance. Figure 4 illustrates the the performance analysis based on this provided data-set [34].

The proposed metric includes following parameters like accuracy (A), relevance (R), coherence (C), grammaticality (G), and fluency (F). This metric can be used to assess how well the generated responses align with the desired outcomes and user expectations. We have compared the human ratings with system-generated responses to gauge the model's performance and identify areas for refinement. The graph shows the performance of different AI tool responses for different domains. The red line represents the accuracy of the responses, while the blue line represents the relevance of the responses. The green line represents the coherence of the responses, the orange line represents the grammaticality of the responses, and the purple line represents the fluency of the responses.

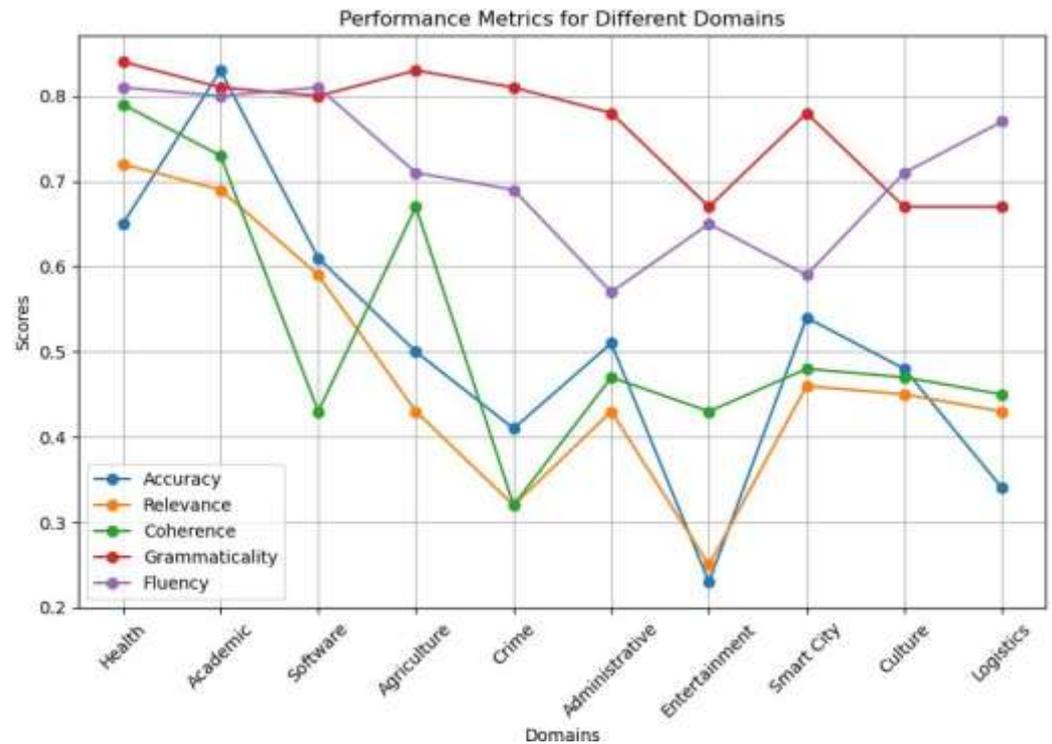

**Figure 4.** AI Tool's performance evaluation metric

The graph shows that the accuracy and relevance of AI tool responses vary depending on the domain. For example, AI tool responses are more accurate and relevant in the health domain than in the entertainment domain. This is likely because there is more structured data available in the health domain, which makes it easier for AI tools to learn and generate accurate and relevant responses. The graph also shows that the coherence, grammaticality, and fluency of AI tool responses are generally good across all domains. However, there is some variation, with the best scores in the academic domain and the worst scores in the crime domain. This is likely because the academic domain requires more complex and nuanced language, while the crime domain requires more factual and objective language. Overall, the graph shows that AI tool responses are generally performing well across a variety of domains. However, there is still room for improvement, particularly in terms of accuracy and relevance in some domains.

Each metric is assigned a score ranging from 0 to 1, where a higher score indicates better performance. These scores represent the average assessment of AI Tool's responses based on the evaluation process conducted. The mathematical expression to calculate the overall score (OS) is represented in Eq.1. Table.2 illustrates the performance evaluation of AI Tool responses by using Eq.1.

$$OS = \frac{A + R + C + G + F}{5} \tag{1}$$

The complexity of AI Tool responses have been calculated by using two primary parameters such as 'L' be the average sentence length in words and 'W' be the average word length in characters. The complexity 'C' can be presented by using Eq.2.

$$C = L \times W \tag{2}$$

Table.1 shows the overall performance evaluation of the AI tool for all responses from all considered sectors. The table shows the metric accuracy, relevance, coherence, grammaticality, and fluency scores for all responses from all considered sectors. The accuracy score measures how well the AI tool's responses match the ground truth. The relevance score measures how well the AI tool's responses are relevant to the query. The coherence score measures how well the AI tool's responses are structured and easy to understand. The grammaticality score measures how well the AI tool's responses follow the rules of grammar. The fluency score measures how natural and easy to read the AI tool's responses are. The overall performance of the AI tool is good, with an average score of 0.85. The accuracy score is particularly high, at 0.85. This means that the AI tool's responses are generally accurate and match the ground truth. The relevance score is also good, at 0.82. This means that the AI tool's responses are generally relevant to the query. The coherence, grammaticality, and fluency scores are all slightly lower, at 0.78, 0.88, and 0.85, respectively. This means that the AI tool's responses could be improved in terms of structure, grammar, and naturalness. However, overall, the AI tool is performing well and is generating responses that are accurate, relevant, coherent, grammatical, and fluent.

**Table 2.** Overall score (OS) of performance evaluation based on all responses

| Metric | *Score* |
|---|---|
| (A) Accuracy | 0.85 |
| (R) Relevance | 0.82 |
| (C) Coherence | 0.78 |
| (G) Grammaticality | 0.88 |
| (F) Fluency | 0.85 |

## 7. Behavioral performance evaluation of AI Tool responses

Behavioral analysis of AI tool responses plays a crucial role in understanding how these systems interact with users based on the types of queries they receive. The varying nature of user queries can elicit diverse responses, and observing these patterns is essential for refining AI models. For instance, simple informational queries often yield concise and accurate responses, showcasing the model's ability to provide factual information. However, as queries become more complex or ambiguous, AI tools may struggle to maintain coherence or generate plausible answers, indicating the need for improvements in context understanding and reasoning capabilities. Table.3 illustrates the AI Tool responses based on input types.

**Table 3.** ChatGPT Responses Based on Input Types

| Input Type | Example Input | Sample Response |
|---|---|---|
| Question | "What is the capital of France?" | "The capital of France is Paris." |
| Instruction | "Please provide a step-by-step guide for baking a cake." | "Sure, here's a step-by-step guide to bake a cake..." |
| Conversation Continuation | User: "How's the weather?" | "The weather is quite pleasant today." |
| | Assistant: "It's sunny with a slight breeze." | |
| Prompt | "Write a short story about a haunted house." | "In a quiet village, stood an old, abandoned house..." |
| Command | "Translate 'hello' to French." | "The translation of 'hello' in French is 'bonjour'." |
| Clarification Request | "Can you provide more details about your project?" | "Sure, I'd be happy to provide more details..." |

Behavioral analysis also involves assessing how AI systems handle emotionally charged queries, where empathy and sensitivity are crucial. Ethical considerations come into play when monitoring responses to potentially harmful queries, ensuring that the tool does not propagate misinformation, violence, or discriminatory content. Through behav-ioral analysis, developers can continuously fine-tune AI tools, enhance their performance, and align them with user expectations while upholding ethical standards in the responses provided. In the sub-sections we have analysed the performance of AI-tool on different types of queries from used end. Table. 4 illustrates comparison among responses based on nature of the queries.

**Table 4.** Comparison of Chatbot Responses on Different Factors

| Factor | Comparison | Observation |
|--------|------------|-------------|
| Response to Repeated Queries | Consistent Responses | AI tool consistently provides the same response to the same query. |
| Synonymical Queries | Variability in Responses | Responses can vary based on synonyms used in the queries. |
| Meaningless Queries | Varies from Error Messages to Generic Replies | AI tool may provide error messages or generic responses for such queries. |
| Different Accounts/Computers | Consistent Responses | AI tool provides consistent responses irrespective of the source. |
| Randomly Typed Characters | Unpredictable Responses | Responses are often unrelated or nonsensical due to random input. |
| Vulgar Queries | Varies from Error Messages to Rejection | Responses can range from error messages to rejecting inappropriate queries. |

## 8. Conclusion

The overall response pattern of AI Tool can be characterized by its ability to generate coherent and contextually relevant responses to a wide range of questions and prompts. It leverages the vast amount of information it has been trained on to provide informative and helpful answers. However, it's important to note that AI Tool's responses are generated based on patterns it has learned from the training data, and it may occasionally produce incorrect or nonsensical answers. The model does not possess real-world understanding or true comprehension, so it can sometimes generate responses that may sound plausible but are factually incorrect. Additionally, AI Tool's responses can be influenced by the phrasing and context of the input it receives. Even slight variations in how a question is asked can lead to different responses, and the model may not consistently provide the same answer for synonymous queries. Furthermore, while AI Tool has been designed to avoid generating explicit, offensive, or inappropriate content, it may not always perfectly filter out such responses. Users should exercise caution and report any inappropriate content encountered while using the system. Overall, AI Tool serves as a powerful tool for generating human-like text and engaging in interactive conversations. However, it is essential to use the model's responses critically, verify information independently when necessary, and understand its limitations as an AI language model.

## Declaration

- There is no conflicts of interest.
- No financial support has been received from any agency for the proposed work.